\documentstyle[aps,psfig,epsfig]{revtex}
\def\be{\begin{equation}}
\def\bea{\begin{eqnarray}}
\def\ee{\end{equation}}
\def\eea{\end{eqnarray}}

\def\ov{\overline}
\def\ra{\rangle}
\def\la{\langle}
\def\r{\right}
\def\l{\left}

\def\d{\delta}

\def\C{{\bf C}}
\def\S{{\bf S}}

\def\t{\tau}

\begin{document}

\title{Neutral evolution of model proteins: diffusion in sequence space
and overdispersion.}
\author{
Ugo Bastolla$^{(a)}$\footnote{Present address: Freie Universit\"at Berlin, 
FB Chemie, Takustr. 6, D-14195 Berlin, Germany}, 
H. Eduardo Roman$^{(b)}$ and 
Michele Vendruscolo$^{(c)}$}
\address{$^{(a)}$HLRZ, Forschungszentrum J\"ulich, D-52425 J\"ulich, Germany\\
$^{(b)}$Dipartimento di Fisica and INFN, Universit\`a di Milano, I-20133
Milano Italy \\
$^{(c)}$Department of Physics of Complex Systems, Weizmann Institute of Science, Rehovot 76100, Israel}

\address{
\centering{
\medskip\em
{}~\\
\begin{minipage}{14cm}
{}~~~
We simulate the evolution of model protein sequences subject to mutations.
A mutation is considered neutral if it conserves 
1) the structure of the ground state, 
2) its thermodynamic stability and 
3) its kinetic accessibility.
All other mutations are considered lethal and are rejected.
We adopt a lattice model, amenable to a reliable solution
of the protein folding problem. 
We prove the existence of extended neutral networks in sequence space -- 
sequences can evolve until their similarity with the starting point 
is almost the same as for random sequences.
Furthermore, we find that the rate of neutral mutations 
has a broad distribution in sequence space. 
Due to this fact, the substitution process is overdispersed 
(the ratio between variance and mean is larger than one). 
This result is in contrast with the simplest model of neutral evolution, 
which assumes a Poisson process for substitutions, 
and in qualitative agreement with biological data. 
{}~\\
{}~\\
\end{minipage}
}}
\maketitle

\section{Introduction}
\label{sec:introduction}

A recent study on the Protein Data Bank (PDB) showed that the 
distribution of pairwise sequence identity between structurally
homologous proteins presents a large Gaussian peak 
at 8.5\% sequence identity, only slightly larger than 
what expected in the 
purely random case\footnote{
The number of amino acid matches obtained by pairing
two random sequences of the same length is given by
the binomial distribution
with $p=1/20$ if one assumes that the 20 amino acids have the
same probability to occur.
For sequences of length $N$ there will be on average
$pN$ identical amino acids, with a variance $Np(1-p)$.
For random sequences, 95\% of pairwise comparisons
yield a sequence identity between 1\% and 9\%.}
(Rost, 1998). This is an interesting result which means
that the structural similarity does not imply sequence similarity.

An intensive computational study on secondary structures of RNA molecules 
(Schuster {\it et al.}, 1994), which is a problem much simpler than
protein folding, and can be studied through efficient and reliable
algorithms, showed that an exponentially
large number of sequences corresponds in average to a single structure,
and the distribution of structures in sequence space 
is quite inhomogeneous (it follows a Zipf law). Sequences folding into 
the most common structures form connected ``neutral networks''
that percolate sequence space. 
These neutral networks directly arise from the non-uniqueness of 
the relation between sequence and structure.

These results are important to understand how 
evolution works at the molecular level.
Kimura (Kimura, 1968; 1983) and Jukes and King (Jukes \& King, 1969)
proposed long time ago that most of the
evolutionary events at the molecular level are non Darwinian. 
They consist in the substitution of one allele with another
one selectively equivalent (neutral evolution). 
The keystone of neutral evolution is the existence of a
``phenotypic threshold''.
The phenotypic threshold is defined in terms
of similarity between structures.
Below a critical value of structural similarity, 
natural selection cannot avoid the spreading of variants of the protein 
in a population through ``diffusion'', 
called by genetists ``genetic drift'' (such threshold
may depend on  the size of the population and on the mutation rate).
Neutral evolution in sequence space conserves the structure of the protein
(thus taking place below the phenotypic threshold)
but can drive to sequence similarities expected for randomly chosen sequences.

The purpose of this paper is to investigate these questions in a
simplified model of protein folding.
Our aim is to show that, as in the case of RNA sequences, 
protein sequence space is traversed by vast neutral networks.

The attempt to study this problem requires some methodological choices.
The first and most severe condition is that, given a sequence of amino acids,
we should be able to identify its native conformation.
This is tantamount of solving the protein folding problem.
We decided to adopt a lattice model of proteins, since,
to date, this is the only case in which protein folding
is routinely solvable. 
This choice restricts drastically
the possibility to represent active sites and structural motifs,
which might play an important role in constraining neutral evolution.

The second choice is how to represent mutations.
We mutate amino acids and not nucleotides.
In nature mutations affect DNA, so that mutations happen with
larger frequency between amino acids whose codes differ by just one
nucleotide. We do not take this fact into account and consider an
uniform probability of mutation for any nucleotide. This
does not change the characteristics of the ``neutral networks'', and it
should not change the relevant features of diffusion in sequence space.
Moreover, we work with sequences of fixed length (number of amino acids) 
$N=36$, considering only point mutations, without insertions and deletions.

The third choice is how to represent the phenotypic threshold.
We introduce a stochastic algorithm which at each time step attempts
to change one amino acid. The mutation is accepted if three conditions are
met: the ground state should be conserved, 
thermodynamically stable and easily accessible.

Since we are limited by computational resources,
we decided to study one particular neutral network.
Namely we followed the evolution from one single common ancestor.
The extent to which this network can be considered paradigmatic
will be commented.

We believe that the features of neutral evolution which are found
in real proteins are universal enough 
to be captured by the present simplified model.
Support to this view comes from a recent 
study by Babajide and coworkers (Babajide {\it et al.}, 1997).
They also found evidence for the presence of neutral networks
in sequence space. Their work is similar in spirit to the present one, 
but significantly different methodologically. They used
a description of the protein structure based on the $C_\alpha$ and 
$C_\beta$ coordinates as taken from the PDB,
and an approximate criterion of fold recognition based on the Z score
(Bowie {\it et al.}, 1991; Casari \& Sippl, 1992).

The paper is organized as follows: 
in Sec. \ref{sec:theory}, we present an outline of the theory
of neutral evolution.
In Sec. \ref{sec:model}, we describe our model
protein and our protocol to simulate neutral evolution. 
In Sec. \ref{sec:drift} 
we illustrate our results concerning the genetic drift. 
In Sec. \ref{sec:pop} we deal with population genetic
considerations and study the dispersion index of the process of
substitutions. 
Sec. \ref{sec:discussion} 
presents an overall discussion, where we discuss the applicability of our
results to the interpretation of existing biological data.

\section{The neutral theory of evolution}
\label{sec:theory}

The theory of evolution explains how different living species evolve
adapting to an ever changing environment. 
The key ingredients of the explanation are reproduction, 
mutation and natural selection, and the complex ecological
interactions that modify the environment.
The pattern of change in time at the phenotypic level is quite
irregular: long periods of stasis are followed by abrupt bursts of activity,
species suddenly appear and disappear and
very large extinction events happen, showing complex features 
(Gould \& Eldredge, 1977; Sol\'e {\it et al.}, 1997).
In contrast to this, evolution at the molecular level seems to be
much more regular. The first studies on this subject showed that
the rate of change in time of the amino acid sequence
of a given protein, which can be estimated from the difference of
homologous sequences of different species, in most cases does not vary
much from one species to another  in the same class, even if the
species compared have rather different population size and
environmental conditions
(Zuckerland \& Pauling, 1962) (on the other hand, the evolution
rate changes drastically from one protein to another).
This property makes each protein a sort of ``molecular clock''
(Zuckerland \& Pauling, 1962; Kimura, 1983; Gillespie, 1991;
Ratner {\it et al.}, 1996), with its own measure of time, and
allows to transform a distance in sequence into a distance in time and
to reconstruct phylogenetic trees from molecular data.

In order to explain this and other features of molecular evolution, as
for example the presence of many polymorphic loci in the genome,
thirty years ago Kimura (Kimura, 1968) and Jukes and King (Jukes \& King,
1969) proposed a new and at that time ``revolutionary''
interpretation of molecular evolution, that was named the neutral
theory of molecular evolution, reviewed in (Kimura, 1983). According to
this theory,
most of the changes in protein sequences happen not because better variants
of the protein are found and spread in the population by natural selection,
but because many mutations do not modify significantly the efficiency 
of the protein, so that natural selection cannot avoid their
spreading through the population by random genetic ``drift''
(``diffusion'' in physical language). 

One decade later, the theory of the molecular quasi-species (Eigen
{\it et al.}, 1989) showed the possible existence of an
evolutionary phase,  at high mutation rate and low selective pressure,
with the diffusive features of neutral evolution: thus this mode of
evolution is expected not only for strictly neutral mutations, but
whenever the damage brought by frequently occurring mutations is
smaller than a threshold depending on the mutation rate.
With regard to proteins, in the framework of their random
heteropolymer model, Shakhnovich and Gutin estimated the probability
of a mutation which does not change the ground state
(Shakhnovich \& Gutin, 1991).
Their conclusion is that this probability has a finite
limit for increasing system size.

There have been however more recent and accurate studies which
question the very existence of molecular clocks, noting that most of
the clocks underwent in some instances ({\it i.e.} in some periods of
the evolution of some lineages) drastic accelerations or
decelerations (Ayala, 1997) so that the usefulness of the clock
hypothesis to reconstruct evolution is very reduced. In some cases,
the adaptive nature of the acceleration of the substitution rate was
clearly demonstrated, like in the case of the changes that
hemoglobin underwent in the time when vertebrates colonized the
earth (Ratner {\it et al.}, 1996).
According to the selectionist interpretation,
these discrepancies of the clocks are such that the neutral hypothesis
should be completely rejected. This dispute produced
a vast literature (see the book by Gillespie (Gillespie, 1991) for a
nice and richly detailed discussion of the selectionist point of
view).

In this work we take a different point of
view, and investigate with a minimal model the possibility of neutral
evolution of proteins. We find that the model supports this
possibility\footnote{However, we stress that the possibility of
  neutral evolution does not imply that neutral evolution did
  occur. In particular, our model describes neutral evolution 
  with the additional hypothesis that the environment remained
  constant. Some selectionist theories, on the other hand, consider a
  rapidly changing environment, so that amino acids substitutions are
  needed to fit the protein to the new environment
  (Gillespie, 1991). While it is undebatable that dramatic climatic and
  ecological changes happen on the time scale of molecular
  evolution, it seems not unreasonable that the homeostatic properties
  of cells protect the cellular environment against such
  changes. Indeed, the most irregularly (thus less ``neutrally'')
  evolving proteins seem to be hormones, which are responsible for 
  intercellular communication, while enzymes evolve in a more regular fashion
  (Gillespie, 1991).
}. Interestingly, we find also that the rate $R$ between the
variance of the number of substitutions and its average value is
larger than 1. Originally, Kimura proposed that the
substitution of amino acids in proteins follows a Poissonian
stochastic process, which implies $R=1$. For most proteins however
real data give $R$ significantly larger than 1, till values between
30 and 50 (Gillespie, 1991) (this problem is produced by the drastic
accelerations of the substitution process
mentioned above). This was considered as a severe evidence against the
neutral theory. Thus it is remarkable that a ``microscopic'' model
which considers only the possibility of neutral evolution leads to
results that reconcile the neutral theory with at least part of the data.

Finally, a remark about terminology: by ``mutation'' we mean the
modification of a triplet of the genetic code in one individual
lineage (by means of environmental damages, errors in the replication
process or other causes). By ``substitution'' we mean the much more
complex process in which an allele dominant in a biological population
is replaced by a new one arising from mutation. We consider here only
individual lineages. Passing from the level of individual lineages
(mutation and reproduction) to the level of a population
(substitution) requires the tools of population genetics. However,
Kimura (Kimura, 1968; Kimura, 1983) showed that the passage is very simple
in the case in which the only mutations occurring are either lethal or
completely neutral for natural selection. In this case, the
substitution rate in the population is

\be \mu_{subs}=\mu x, \label{kim} \ee
where $\mu$ is the ``microscopic'' mutation rate (at the level of the
single lineage) and $x$ is the fraction of mutations which are
selectively neutral (see also (Bastolla \& Peliti, 1991)). Note that the rate of
evolution does not depend on the size of the population, whereas it
does if positive natural selection plays a role.

Under these assumptions, the model of neutral evolution of single
lineages that we present here is equivalent to a model of evolution at
the level of a population, and thus it can be in principle compared to
biological data.

\section{Model of neutral evolution in proteins}
\label{sec:model}

In this section we give a detailed description of the lattice model
we used to represent proteins and of the algorithm we introduced
to study evolution in sequence space.

\subsection{Definition of the model proteins}
In our model, a protein configuration is represented by
a self avoiding walk on the simple cubic lattice. 
Each of the $N$ occupied sites represents an amino acid, chosen
among the 20 possible ones. We represent a configuration of the protein 
by its contact map $\C$. This is a $N\times N$ matrix whose element
$C_{ij}$ is 1 if residues $i$ and $j$ are nearest neighbors
on the lattice (but not along the chain) and 0 otherwise. Note that the
correspondence between contact maps and configurations is not unique, and
the smaller is the ``density of contacts'' $N_c/N$ of a contact map,
where $N_c=\sum_{i<j}C_{ij}$ is the number of contacts,
the larger is the number of configurations to which 
it corresponds (Vendruscolo {\it et al.}, 1998).

We denote a sequence of length $N$ by $\S =\{s_1,\ldots,s_N\}$.
An energy $E(\S,\C)$ is assigned to sequence $\S$ on a configuration whose
contact map is $\C$
\be 
E(\S,\C)=\sum_{i<j}^{1,N}C_{ij} U(s_i,s_j) \,.
\ee
where $U(a,b)$ is a $20\times 20$ symmetric interaction matrix, which gives
the energy gain obtained when amino acids of species $a$ and $b$ are
brought into contact.
We use a matrix $U(a,b)$ derived from the Miyazawa-Jernigan interaction 
matrix (Miyazawa \& Jernigan, 1985).
In our model, we identify the ground state structure as the native structure.

We now define two measures of similarity respectively
in sequence space and in structure space.
The standard measure of similarity between two sequences $\S$ and $\S'$
of the same length $N$ is the Hamming distance, which counts the number of
amino acids which are different:
\be
D(\S,\S') = \sum_{i=1}^N \left[ 1 - \delta (s_i,s'_i) \right ] \; ,
\ee
where $\d$ is the Kronecker symbol and $s_i$ takes 20 different values.
We also consider the overlap $Q(\S,\S')$,
\be 
Q(\S,\S')={1\over N}\sum_{i=1}^N \d(s_i,s'_i) \;,
\label{eq:dh}
\ee
which is equal to 1 minus the normalized Hamming distance $D/N$. 

In sequence space, we introduce also two alternative
measures of similarity,
$D_{HP}(\S,\S')$ to measure differences in hydrophobicity.
and  $D_U(\S,\S')$
to measure the difference in the native interactions.
To define $D_{HP}(\S,\S')$ we transform every sequence into
a sequence of binary symbols, either H or P, 
according to the hydrophobicity of the residue.
We consider 8 hydrophobic amino-acids and 12 polar ones. 
The definitions of $D_{HP}(\S,\S')$ and of the related overlap
$Q_{HP}(\S,\S')$ 
are analogous to those of $D$ and $Q$, 
where now $s_i$ can take only two values.
The distance $D_U(\S,\S')$ is given by the quadratic difference of 
the native interactions in two different sequences:
\be 
D_U(\S,\S')=\sum_{i<j} \C^*_{ij} \left[ U(S_i,S_j)-U(S'_i,S'_j)
\right]^2 \;, \label{DU} \ee
where $\C^*$ is the native contact map.

We measure similarity in configuration space 
by the overlap $q(\C,\C^\prime)$ between contact maps $\C$ and $\C'$
\be 
q(\C,\C^\prime)={1\over N_c^*}\sum_{i<j}C_{ij} C^\prime_{ij}\;,
 \ee
where $N_c^*$ is the maximal between $N_c$ and $N_c'$, the
number of contacts respectively of two contact maps $\C$ and $\C^\prime$.
Two maps are identical if and only if $q=1$.

\begin{figure}
\centerline{\psfig{file=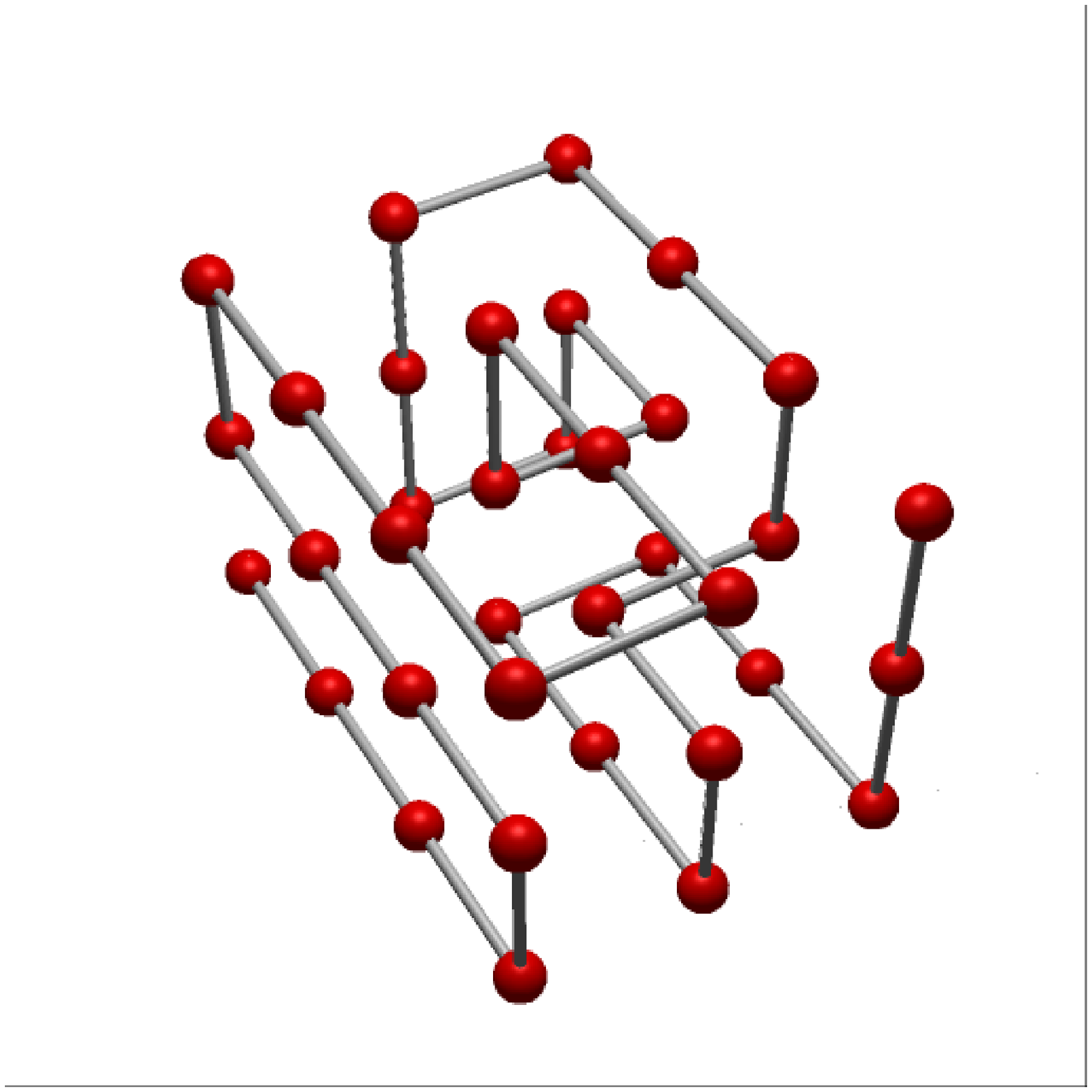,width=7cm,angle=0}}
\caption{The ``native state'' of our model protein}
\label{fig:conf}
\end{figure}

We consider a sequence with $N=36$ 
and a contact map $\C^*$ which has the highest possible number of
contacts for this chain length, $N_c=40$. (See Fig. \ref{fig:conf}).
In this case the contact map defines uniquely the configuration
of the system (apart from trivial symmetries). 

The contact map $C^*$ was studied by Shakhnovich and coworkers 
in a computer experiment of inverse folding (Abkevich {\it et al.}, 1994).
They designed a sequence $\S^*$ with ground state on $\C^*$ using the
procedure of (Shakhnovich \& Gutin, 1993),
showing that $\S^*$ has good properties of kinetic foldability and
thermodynamic stability at the temperature where the folding is fastest.
The lower part of the energy landscape of this
sequence is remarkably smooth:
all the structures with low energy have a high overlap $q_0$ 
with the ground state. The lowest energy of configurations with a
fixed value of $q_0$ decreases regularly as $q_0$ approaches one. 
This correlated energy landscape, 
reminiscent of the ``funnel'' paradigm (Bringelson \& Wolynes, 1987), 
is the reason of the good folding properties of the sequence, 
which is very different from a random one. 
In (Tiana {\it et al.}, 1998) it was shown that the same sequence is also very
stable against mutations. It was estimated that
about 70\% of the point mutations performed on $\S^*$
result in a new sequences with exactly the same ground 
state and good folding properties. 
Thus energy minimization makes $\C^*$ stable not only in structure
space, but also in sequence space.

We note that $\C^*$ is a rather atypical structure 
for the interaction parameters that we choose: 
since $U(a,b)$ has average value zero and variance $0.3$, we would
expect open structures to be energetically favored. 
Indeed, typical random sequences with $N=36$ and contact interactions
whose average vanishes have a ground state with approximately 30-32 
contacts (Bastolla, unpublished result), being thus less than
maximally compact. 

\subsection{Definition of the model of evolution}

We consider a reference contact map $\C^*$ as the biologically 
active native structure, and we set as starting point of 
neutral evolution a sequence $\S^*$ which folds onto $\C^*$. 

The simulated molecular evolution is realized 
by the following iterative procedure
\begin{enumerate}
\item
At $t=0$ we start from $\S(0)=\S^*$.
\item
At time step $t$ we mutate at random one amino acid of $\S(t-1)$,
producing a new sequence $\S'(t)$. 
\item
We submit the new sequence to selection according to the criteria
specified below.
If the sequence survives then $\S(t)=\S'(t)$,
otherwise we restore $\S(t)=\S(t-1)$.
\end{enumerate}
The selection of sequences is governed by the following conditions:


\begin{description}
\item{\em Conservation of the phenotype}.

The ground state $\C$ of $\S'(t)$ must have an overlap with $\C^*$
not smaller than a given ``phenotypic threshold'' $q_{\rm thr}$.
\be 
q(\C^*,\C) \geq  q_{\rm thr} \;, 
\ee

In our calculations we imposed strict conservation of the phenotype
by setting $q_{\rm thr}=1$.

\item{\em Thermodynamic stability}.

We define thermodynamic stability through the condition
\be 
\l\la q(\C^*,\C)\r\ra \geq \la q\ra_{\rm thr} \;, 
\ee
where $\la \cdot\ra$ represent a Boltzmann average at the temperature $T$ 
of the simulation and $\la q\ra_{\rm thr}$ is a fixed parameter.
This condition implies that all the thermodynamically relevant states
have an overlap greater than $\la q\ra_{\rm thr}$ with the target state.

\item{\em Kinetic accessibility}.

We use the PERM method (Grassberger, 1997; Fraeunkron {\it et al.}, 1998;
Bastolla {\it et al.}, 1998), a new Monte Carlo algorithm
which is particularly successful in finding the 
ground state of lattice polymers.  
For every sequence $\S$ to be tested, we run $\t$ iterations of the
algorithm. If the lowest energy structure found at this point does not
coincide with $\C^*$, we discard the sequence because the kinetic
accessibility condition is violated. Otherwise, we continue to run the
algorithm for a time $\t$ to check that no lower energy structure is
found. If this condition is met, we start again the algorithm and run
it for a time $2\t$. Only if $\C^*$ is once again found as the lowest
energy structure we conclude that the ground state of $\S$ really
coincides with $\C^*$. Note that there is no bias towards $\C^*$ in
our Monte Carlo algorithm.

\end{description}

We never found in the second independent run of the MC
algorithm a structure with lower energy than the putative
ground-state $\C^*$. This fact encourages us in believing that the
algorithm was really effective in finding the ground state. Another
support to this conclusion comes from the fact that all of the
selected sequences had a remarkably correlated energy landscape, which
makes the search for the ground state easier.
On the other hand, whenever a sequence was rejected, 
we are not sure whether we were able to determine its ground
state. The difference is due to two reasons: first, we run rejected
sequences for a shorter time ($\t$ instead of $2\t$ is
the rejection is made at the first decision stage, 1 run instead of 2
independent ones if it is made at the second stage). Second, rejected
sequences have typically a much less correlated energy landscape, so
that it is expected that the determination of the ground state is more
difficult. Nevertheless, we shall discuss at the end also
data about rejected sequences, since they are interesting and
refer to a very large number of sequences, even if they are
individually not completely reliable. 

\vspace{.5cm}
The 3 conditions for the acceptance of a mutation enforce
the conservation of biological activity of the new sequence. We have
to stress that conservation of the fold is not a necessary condition
for selective neutrality in the real world, and it is not even
sufficient, since the active site has also to be conserved and the
environment has to remain reasonably stable. Thus with our model we
can represent just the neutral evolution of the part of the chain not
involved in chemical activity, in a stable chemical
environment. Nevertheless, we think that these limitations 
do not prevent from the applicability of the model to real molecular
evolution.

\subsection{Genetic drift}
\label{sec:remarks}

For a given sequence $\S$ of $N$ amino acids,
we define the rate of neutral mutation $x$ as the fraction 
of acceptable non-synonymous mutations
\be
x(\S)= \frac{1}{20N}\sum_{i=1}^{N}\sum_{\alpha\neq s_i}^{1,20}
\chi_i^\alpha \; , \label{eq:x}
\ee
where $\chi_i^\alpha=1$ if assigning the amino acid of species $\alpha$
at position $i$ on the sequence does not change the native state
and 0 otherwise. In Kimura's ordinary neutral theory (Kimura, 1983) it
is assumed that $x(\S)=x$ is indeed independent of the sequence. With this
hypothesis, the evolution of the overlap from the
starting sequence $\ov{Q(t)}=\ov{Q(\S(0),\S(t))}$ (the overbar denotes an
average with respect to the mutation process) is given by
\be 
\ov{Q(t)} \approx {1\over 20}+\l(1-{1\over 20}\r) {\rm e}^{-xt/N}\;.
\label{eq:xexp}
\ee

Since every mutation is either neutral or lethal, the evolution of the
wild-type genome of the population coincides with that of a single
reproductive line (Kimura, 1983), and we may interpret different realizations
of our evolutionary process as different `species' originating from a
common ancestor. Thus $\mu_{subs}=x/N$
is the substitution rate both of a single lineage and of the
population, and the time $t$ represents in our model the number of
mutational events. As we shall see, however, the hypothesis of constance of 
$x(\S)$ is not in agreement with the results of the simulations.

\subsection{Neutral set}

We define the neutral network $\omega(\C^*,\S^*)$ as follows:
\begin{quote}
$\omega(\C^*,\S^*)$ is formed by all the sequences $\S$ 
that have $\C^*$ as their 
ground state {\em and} are connected to $\S^*$ 
through a {\it neutral path} (a path in sequence space which satisfies
the three selection criteria given above).
\end{quote}

We then define the neutral set $\Omega(\C^*)$ as
\begin{equation}
\Omega(\C^*) = \cup_\mu \omega(\C^*,\S_\mu^*)
\end{equation}
with $\mu$ running over different sequences $\S_\mu^*$ with ground state
on $\C^*$ but not connected by neutral paths. It is possible that this set is
larger than the neutral network of a single sequence, $\omega(\C^*,\S^*)$.
In this case it would be possible in principle to distinguish between
convergent (different ancestors, same final structure) and divergent 
(same ancestor) evolution: if the evolution of two homologous proteins takes
place in two disjoint
neutral networks, it must be convergent. 

We studied only one starting
sequence $\S^*$, thus this work is concerned only with divergent evolution.
However, since the neutral networks that we studied is so spread that typical
pairs of sequences have a Hamming distance comparable with that of random
sequences, it turns out that it is not possible to distinguish between
divergent and convergent evolution on the basis of the sequence homology
alone. Under this respect, our results are equivalent to those of Rost
(Rost, 1997).

\section{Results}
\label{sec:drift}


\subsection{Evolutionary drift}
\label{sec:distance}

We generated 8 trajectories with the following values 
of the selection parameters: 
$q_{\rm thr}=1$,
$\la q\ra_{\rm thr}=0.9$, $T=0.16$, $\t=1.6 \cdot 10^5$.
In each trajectory, a number of sequences 
ranging from 875 to 2709 were tested. 
The CPU time needed for a single trajectory 
was about 4 to 5 weeks on Sun Ultra stations.

Different realizations differentiated greatly
from the ``common ancestor'' $\S(0)$ and from each other. 
We give three measures of such differentiation.
1) The average Hamming distance between the final points of the 8
evolutionary trajectories is $D_H=30.2$, which is
only 12\% smaller than the random value $D_H^{ran}=34.2$. 
2) The maximum distances from the starting point in the 8 trajectories are
respectively 26, 28, 27, 27, 17, 33, 25, and 23. 
3) The maximum distance between sequences 
in different trajectories ranges from 27 to 35. 
We note that these values were still increasing as a function
of the length of the simulations when we had to stop them, so that it
is possible that their asymptotic values for very long trajectories
are compatible with those of random samples of sequences. Moreover, the
trajectories were run for different mutational times. 

\begin{figure}
 \centerline{\psfig{file=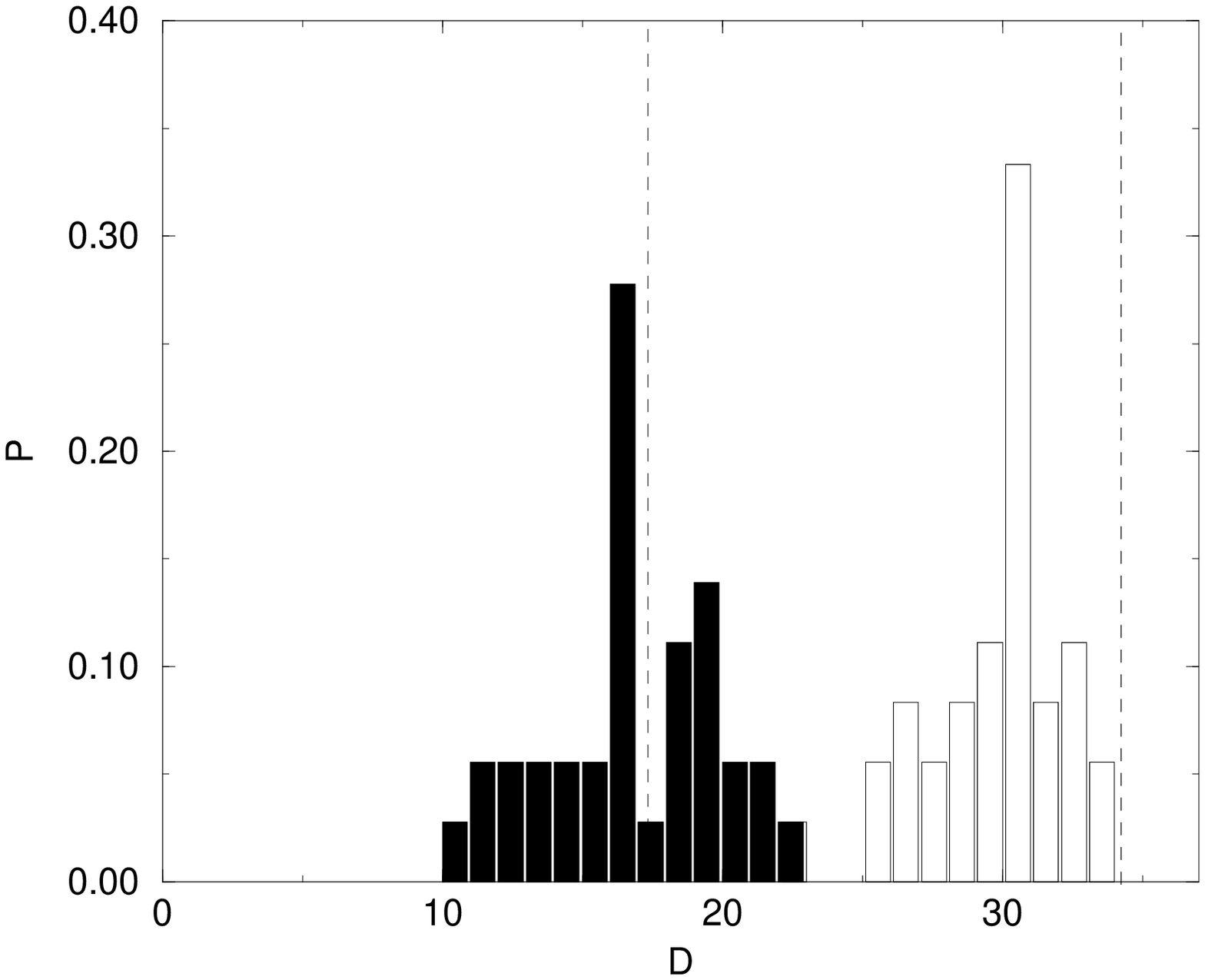,height=7cm,angle=0}
      \psfig{file=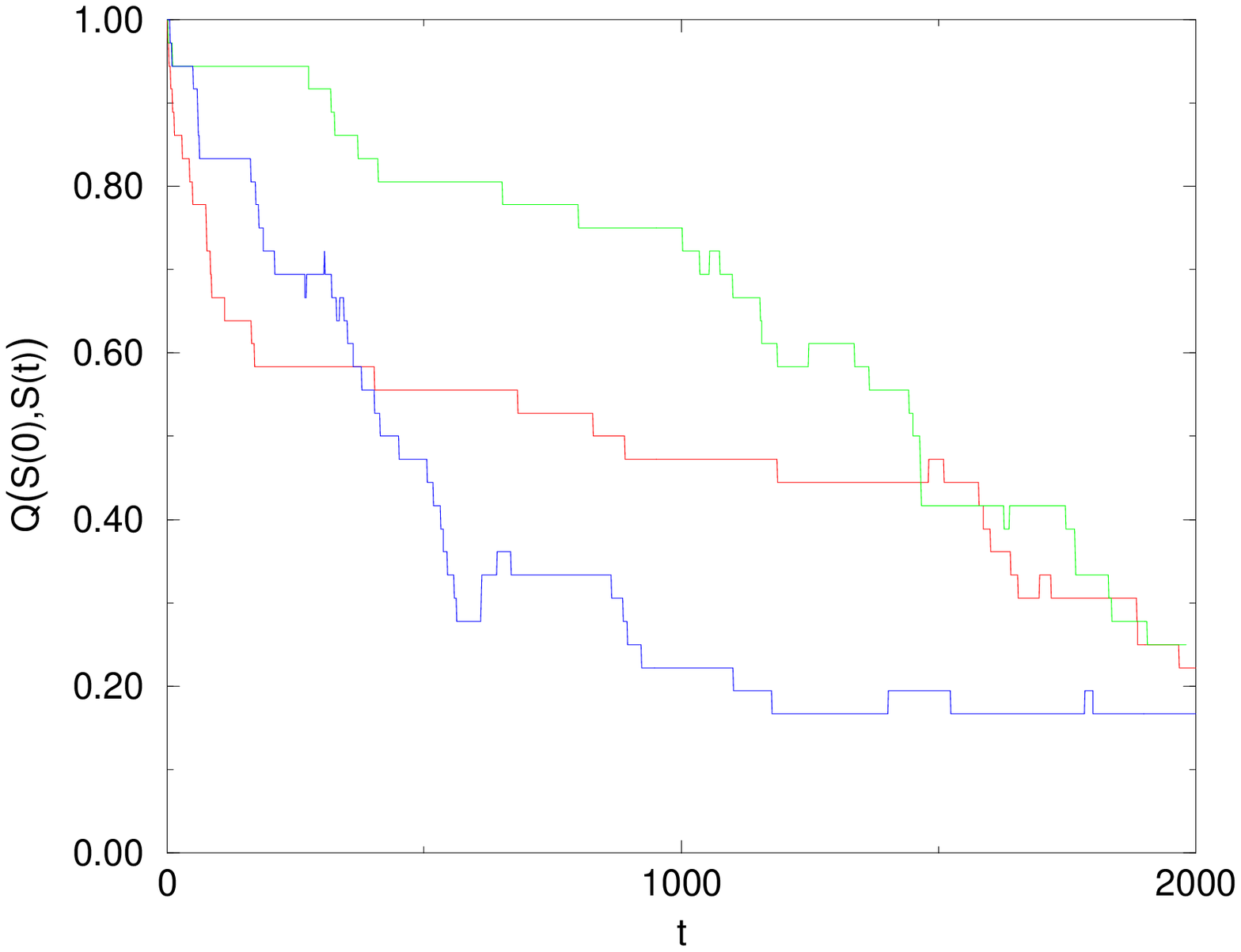,height=7.0cm,angle=0}}
\caption{(a): Histograms of the Hamming distances between the end points of
  independent trajectories with the full 20 amino-acid alphabet
  (white) and with a reduced 2 amino-acid alphabet (black). The dashed lines
  indicate the most probable values of the distance for random pairs (b): Decay
  of the overlap with the starting sequence for 3 independent
  trajectories.}
\label{fig:endpoints}
\end{figure}

Fig. \ref{fig:endpoints}a shows the distribution of the distance among the 
end points of independent trajectories,  both with the full 20 letters
alphabet (white bars) and with the coarse-grained hydrophobic
representation (black bars). In the latter, the average value is
$D_{HP}=16.3$, not far from the random value $D_{HP}^{ran}=17.3$, and 
the variance is $V_{HP}=6.0$, compatible with $V_{HP}=D_{HP}(1-D_{HP})/N$.
In both cases, the typical values for random sequences are larger than
the typical values we found, but this could be due to the
fact that our simulations did not last a time long enough to
equilibrate. It is surprising 
that also the hydrophobic distance is close to that expected for random
sequences. This is probably an effect of the short length of our sequences
whose ground states have at most
2 residues in the core, while all other ones are
at the surface. The two residues at the core are the most conserved
during the evolution (see next subsection). However, we observe that
even one of these residues could be replaced in some instances with
hydrophilic residues.

In Fig. \ref{fig:endpoints}b we show the relaxation of the overlap 
$Q(\S(0),\S(t))$ for three trajectories.
The differences from one realization to another are quite large, 
but at large times the trajectories show the tendency to converge together.

\begin{figure}
\centerline{\psfig{file=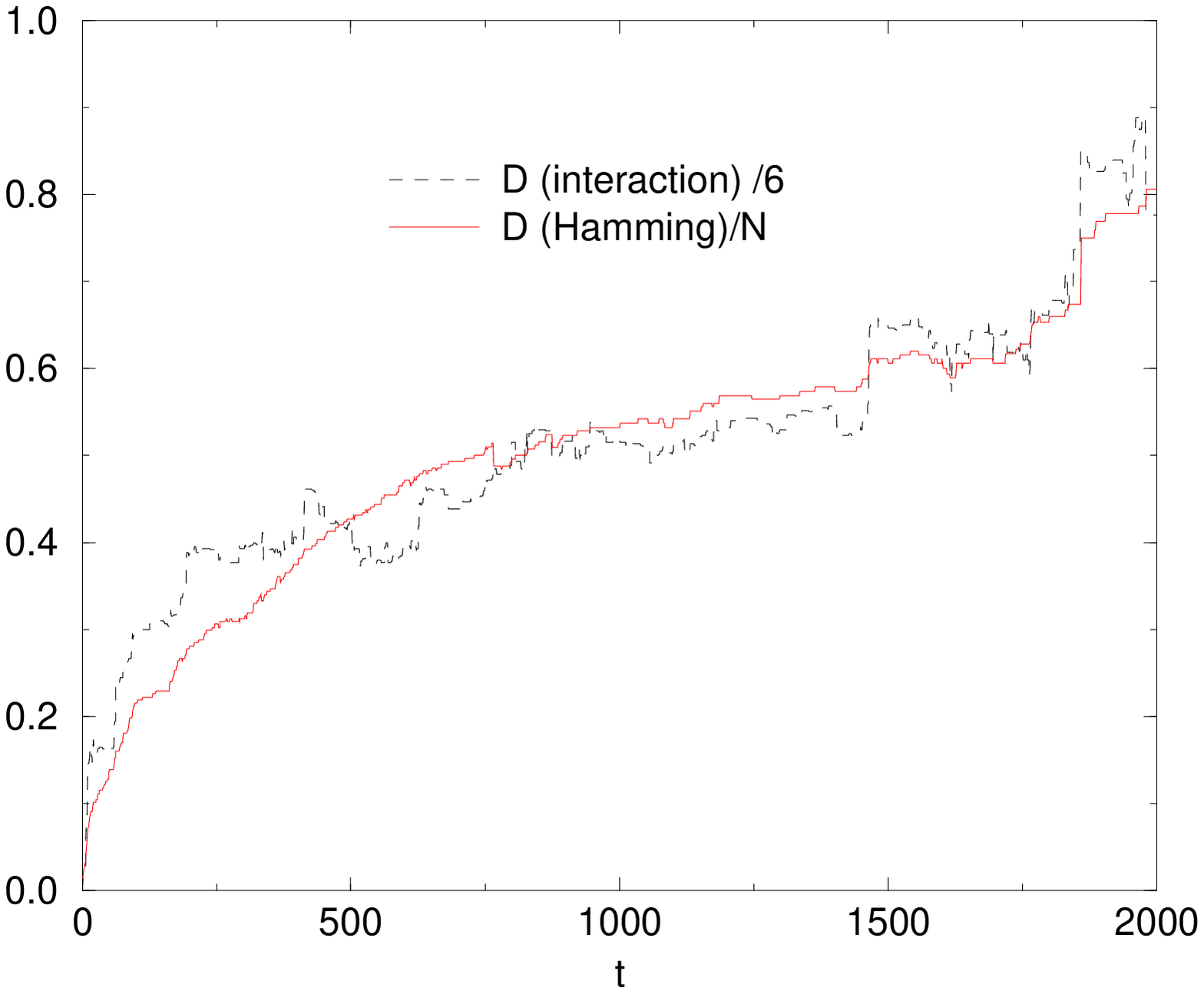,height=7.0cm,angle=0}
            \psfig{file=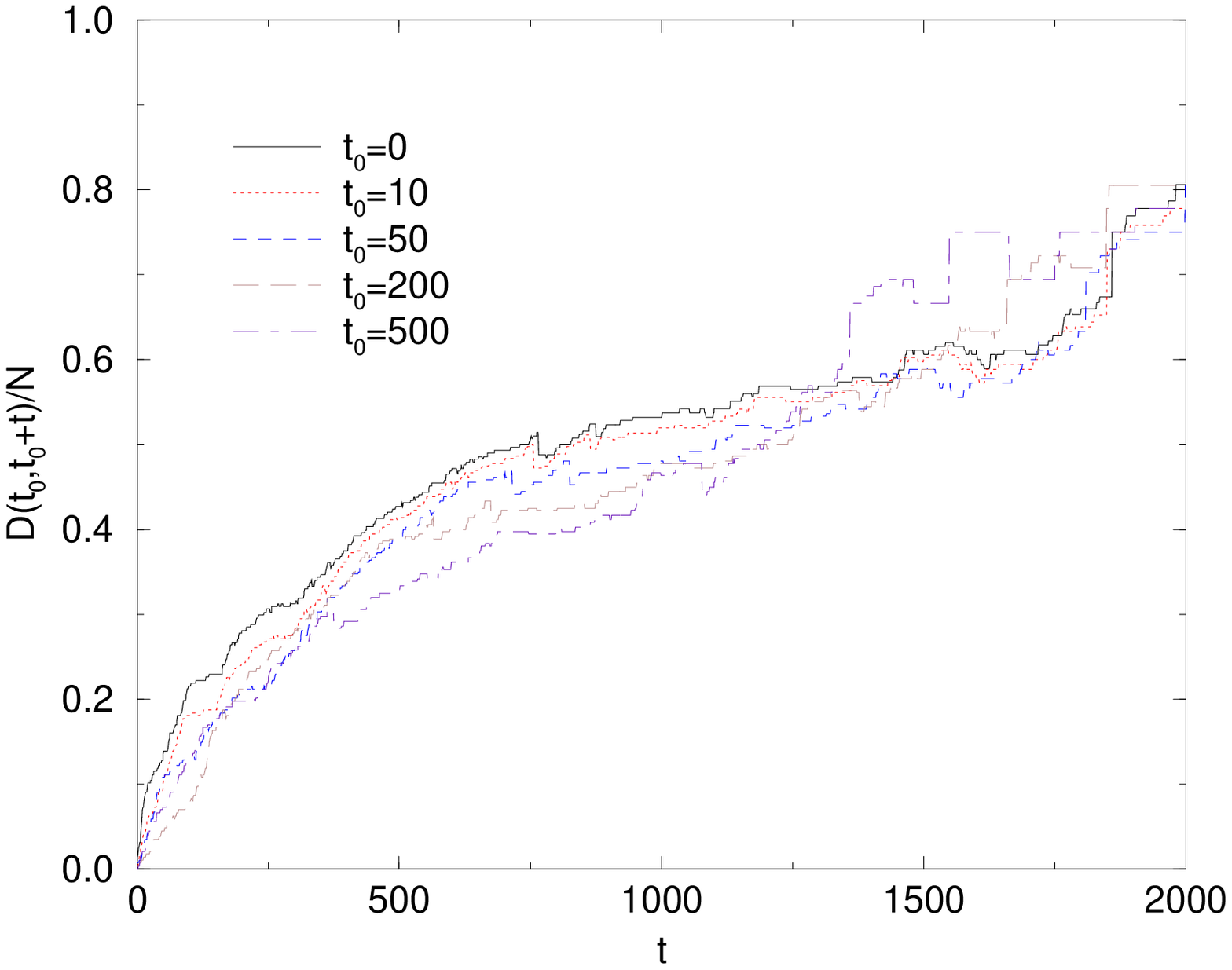,height=7cm,angle=0}}
\caption{(a): Interaction distance (quadratic
  difference of the native interactions) from $\S(0)$ compared with the
  Hamming distance. (b) Distance averaged over different trajectories 
from different sequences $\S(t_0)$.}
\label{fig:distance}
\end{figure}

In Fig. \ref{fig:distance}a we show the Hamming distance
$D(\S(0),\S(t))/N$ averaged over 8 trajectories and we compare it with the
interaction distance $D_U/6$ (Eq. \ref{DU}).
The temporal behavior is very similar in the two cases, although the
maximum value reached in the case of $D_U$ is 5.6, roughly half of the naive
expectation for random sequences given by the number of native 
contacts (40) multiplied by the variance of the interaction matrix (0.3).
In Fig. \ref{fig:distance}b we show $D(t_0,t)/N=1-Q(\S(t_0),\S(t_0+t))$,
averaged over all the trajectories generated. There is a systematic
dependence on $t_0$ at small $t$: the larger is $t_0$, the slower is
the initial relaxation.  We believe that this result is due to the fact that
$\S(0)$ is quite a peculiar starting point, much more stable with
respect to point mutations than other sequences in
$\omega(\C,\S(0))$. Thus, as the trajectories go further away from
$\S(0)$, the rate of accepted mutations decreases. 
On the other hand, the different curves meet again at large
$t$. This could be due to the fact that at large $t$ a large portion of
sequence space has been explored, and the rate of accepted mutations has been
averaged over this large region.

The relaxation in Fig. \ref{fig:distance}a is not exponential, 
as it would be if the probability $x$ of a neutral mutation were
almost constant for all sequences. In fact, as we shall see, the
quantity $x$ defined in Eq. \ref{eq:x} has a very broad distribution in
sequence space, and the simple expectation (Eq. \ref{eq:xexp}) does not hold.

\subsection{Rigidity}

\begin{figure}
\centerline{\psfig{figure=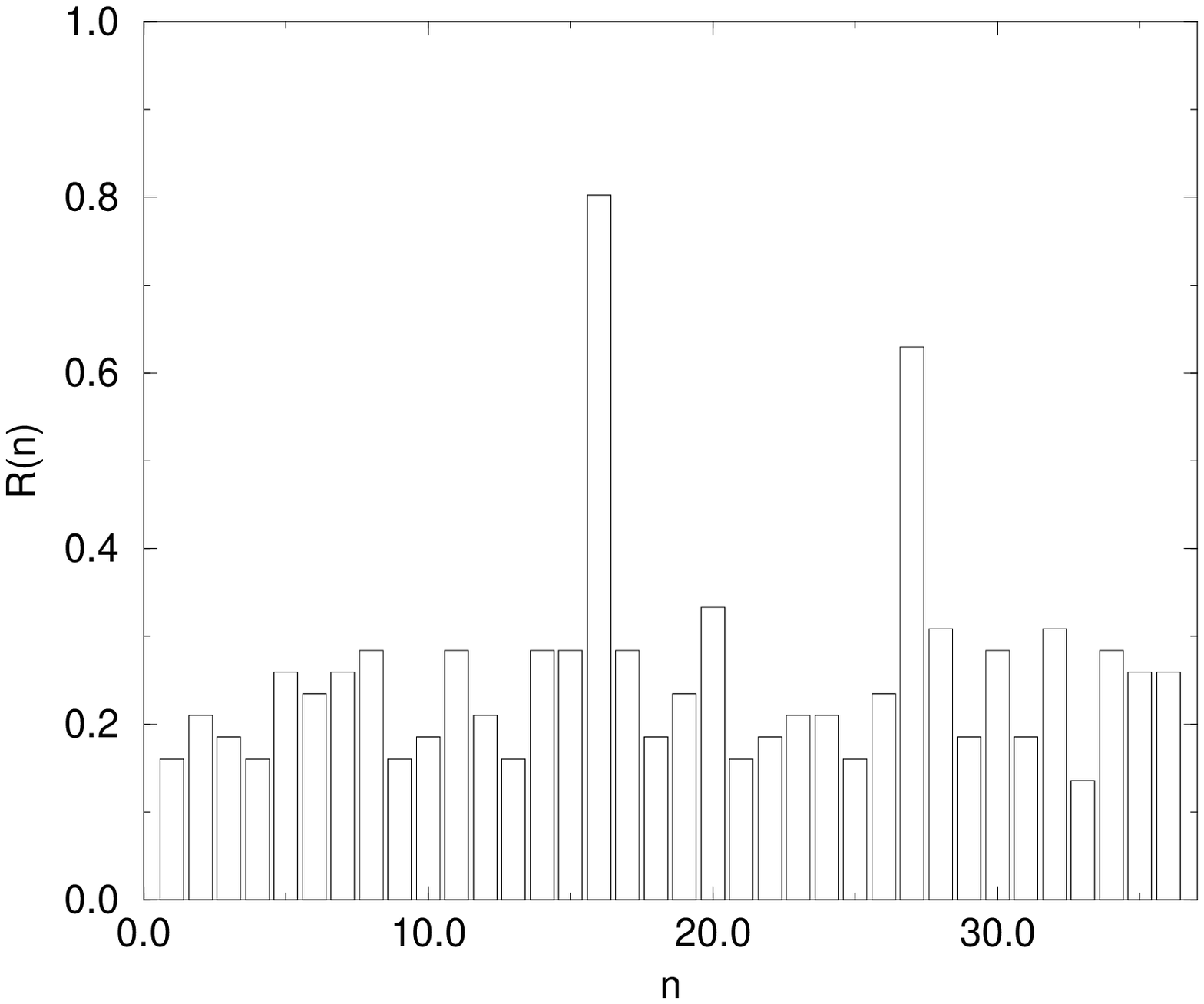,height=7.0cm,angle=0}
            \psfig{figure=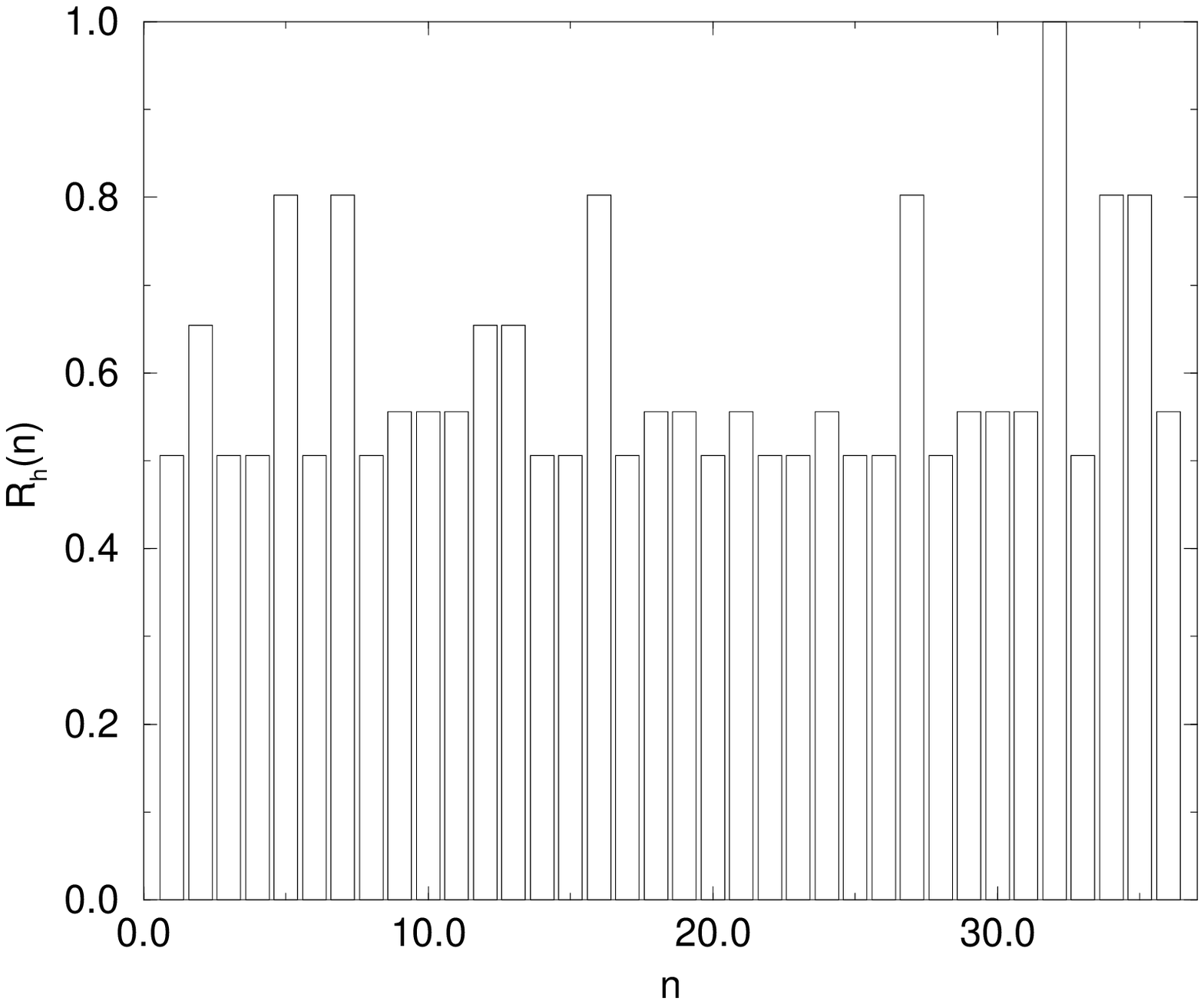,height=7.0cm,angle=0}}
\caption{Rigidity at different positions (see text) for (a) the full 20
  amino-acids alphabet and (b) a reduced HP alphabet. The peaks in (a)
  correspond to the two core positions.}
\label{fig:rigidity}
\end{figure}

We define the rigidity $R_i$ as a measure of the degree of conservation 
of residue $i$:
\be 
R_i=\sum_a P_i^2(a)  \;,
\ee
where $P_i(a)$ is the probability to find the amino-acid $a$ at position $i$.
$P_i(a)$ is estimated from the end points of the 8 neutral paths we generated.
$R_i=1$ if the amino-acid at position $i$ is never changed, 
while $R_i=1/20$ if it is completely random. We show in
Fig. \ref{fig:rigidity} that all the amino acids could be changed at
least once, even if the value of $R$ is typically 3 to 5 times larger than
for a random distribution, and some sites are very stable. 
Not surprisingly, the two most stable sites are in the `core' of
the protein. One of them has already been found to be
particularly sensitive to mutations concerning $\S(0)$ (a red site, in
the terminology of Tiana {\it et al.}). It
is remarkable that, even if the second core site was not very
sensitive to mutations in $\S(0)$ (it was classified as a yellow site), we
find that it is strongly conserved in the overall evolution. 
Fig. \ref{fig:rigidity}a shows the rigidity for the 20 letter alphabet.

In Fig. \ref{fig:rigidity}b we show the rigidity for the coarse-grained HP
alphabet. This is of course larger than in the case of 20 letters.
Many residues have a rigidity compatible with the random value $R_{HP}=
0.52$, and the hydrophobicity of every residue was changed at least once 
in the course of evolution (even if amino acid 32 is always polar in
all of the 8 final sequences). At first glance these results seem surprising,
since one would expect that the hydrophobic pattern should be more conserved 
during evolution than we actually found. However, our model protein is quite 
small, and its hydrophobic core is constituted by only 2 sites, whose
rigidity is much larger than random, and it is not strange that the
hydrophobic pattern of most of the other residues is close to a
random one.

\subsection{Neutral Mutation Rate}

The simplest measure of the neutral mutation rate $x$ (Eq. \ref{eq:x})
is obtained by computing the frequency of neutral mutations over all the
non-synonymous mutations proposed. In this way we found
$\ov{x}\approx 0.06$ (the overline represents an average over
the mutational process). However, this quantity alone is not enough to
characterize $x$, which fluctuates strongly in sequence space. For
instance, it was estimated by one of us and coworkers
(Tiana {\it et al.}, 1998) that
$x(\S^*)\approx 0.7$, where $\S^*$ is the starting point of all our
evolutionary trajectories.

In order to see whether $x(\S)$ has some structure in sequence space,
we divided the sequences $\S\in\omega(\C^*,\S(0))$ in every
trajectory in groups of 100 sequences, as long as they are generated, 
and we stored the average fraction of successful mutations in each group, 
$x(r)$ (where $r$ labels the sequences in the order 
in which they are generated). It appears that the first groups of
sequences have a rather high value of $x$, but this value quickly decreases,
and then  $x(r)$ seems to fluctuate more or less randomly.

\begin{figure}
  \centerline{\psfig{file=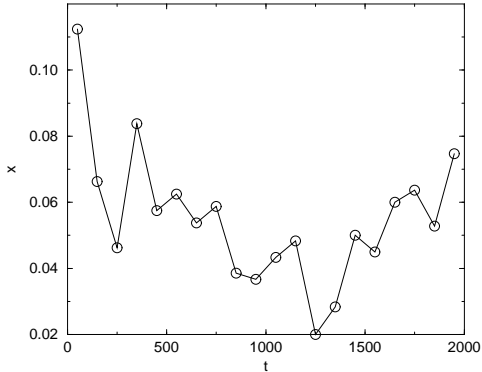,width=7cm,angle=0}}
\caption{Variation of $x(S)$ in sequence space.}
\label{fig:xt}
\end{figure}

We also measured indirectly the distribution of $x(\S)$ in sequence space
from the distribution of the ``trapping'' time $\t_j(\S)$ 
that a trajectory spends on sequence $\S$. The average value of the
trapping time is inversely proportional to the neutral mutation rate 
(we neglect in this argument
the randomness given by the error in evaluating whether a sequence
belongs to the neutral set: in particular, the conditions of fast
folding and of thermodynamic stability are subject to considerable
evaluation errors): 
\be
\ov{\t_j(\S)}=\frac{1}{x(\S)} \; ,
\ee
where the bar denotes average over different attempts to mutate sequence
$\S$. These attempts are unsuccessful with probability $1-x(\S)$, so that
the probability that the first successful mutation is met at trial number
$\t$ is given by the geometric distribution,

\be P_x(\t)=x(1-x)^{\t-1}\: . \ee

\begin{figure}
  \centerline{\psfig{file=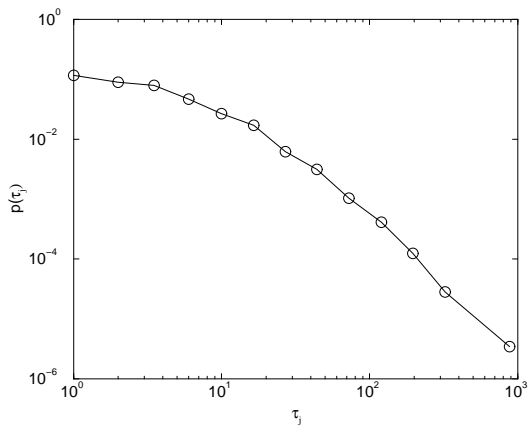,width=7cm,angle=0}}
\caption{Distribution of the trapping time $\tau_j$.}
\label{fig:trapping}
\end{figure}

Averaging in sequence space, we get
\be 
\l[P(\t)\r]=\int_0^1 dx~p(x)\l(x\over 1-x\r)(1-x)^\t \; \label{eq:Ptau},
\ee
where $\l[\cdot \r]$ denotes an average over sequences belonging to 
the neutral network $\omega(\C^*,\S(0))$.
We found that the distribution of $\t_j$ is very broad. It seems to be
broader than an exponential (see Fig. \ref{fig:trapping}), thus, even if we
cannot invert Eq. \ref{eq:Ptau}, we expect that the distribution of the
neutral mutation rate $x$ is also broader than exponential.

We measured also the correlations of $\t_j$ in sequence space,
\be 
C_\t(l)={\ov{\t_j(r)\t_j(r+1)}\over \ov{\t_j(r)}^2}-1 \;. 
\ee 

There is a positive correlation after one step in sequence space,
$C_\t(1)\approx 0.98$, but after few steps the correlation vanishes
(data not shown).

\subsection{Rejected sequences}

As the last point, we want to report briefly results regarding the
sequences that were rejected by our selection algorithm. More details
about this point and about properties of selected sequences will be given
in a forecoming publication. As we said, results concerning
rejected sequences are not completely reliable, since in this case the
identification of the ground state is only tentative. However, they
represent about $94\%$ of the $12,000$ sequences that we
generated, and the statistical properties of this large set are
interesting and qualitatively clear. The most interesting observation
concerns the overlap $q_0=q(\C^*,\C_0)$ 
measuring the similarity between the ground states $\C_0$
of our sequences and the target structure $\C^*$. This has a bimodal
distribution, with a high peak at $q_0=1$ (more than 16\% of the
rejected sequences), corresponding to sequences that have ground state
on $\C^*$ but do not fulfill either the condition of thermodynamical
stability or that of fast folding. This peak has a sudden drop
and then it decreases slowly at decreasing $q_0$. At $q_0=0.32$ a new peak
is present, related to ground states which have less contacts than $\C^*$
(the typical value is 34 instead of 40), but more than it is expected
in the ground state of random sequences. These structures are more
similar to typical low energy structures than to the target state. Thus,
even if a large fraction of mutations conserves the ground state, the
majority of them produces very large structural changes.

The energy of the ground state is strongly correlated both to 
the similarity $q_0$ to the target
and to the number of contacts $N_c^0$ (in fact, the latter two quantities are
strictly related).  Only in few cases we found
sequences with very low ground states that are unrelated to the target: 
for instance, in two cases we found energies lower than -17 with $q_0$ 
as low as $.5$ and $.35$ respectively. From this observation we speculate
that it is unlikely (although not excluded) that two neutral sets
of stable and unrelated structures can be close to each other.
In this sense, our result is related to those of (Li {\it et al.}, 1998).

\section{Overdispersion and population genetics considerations}
\label{sec:pop}

In the previous sections, we described the evolution of a single lineage
of a protein, subject only to lethal and neutral mutations. As we
mentioned in Sec. \ref{sec:theory}, under these hypothesis the evolution
at the level of the population happens with the same rate as the
evolution of a single lineage. Thus we can interpret our 8
evolutionary trajectories as 8 species differentiating from a common
ancestor (star phylogeny), and compare our data to real evolutionary
rates.

In order to do this comparison, we have to further elaborate on the
model for mutations. Time in the model is measured as the number of
mutation events, and we have to relate this number to real time. The
simplest possibility is to assume that the number of mutations
in the geological
time $T$ is a Poissonian variable with average value $\mu T$. This
assumption is similar to Kimura's one, but in his model
the fraction of neutral mutations $x$ is considered constant
throughout the evolution, while our results show that this quantity
is strongly fluctuating.

Thus we simulate the evolution of a star phylogeny by extracting
8 Poissonian variables with average value $\mu T$, $t_1\cdots t_8$. The
number $S_i(t_i)$ of substitutions fixed after $t_i$ mutational events
in the $i-$th trajectory is then interpreted as the number of
substitutions in the species $i$. 

\begin{figure}
\centerline{\psfig{figure=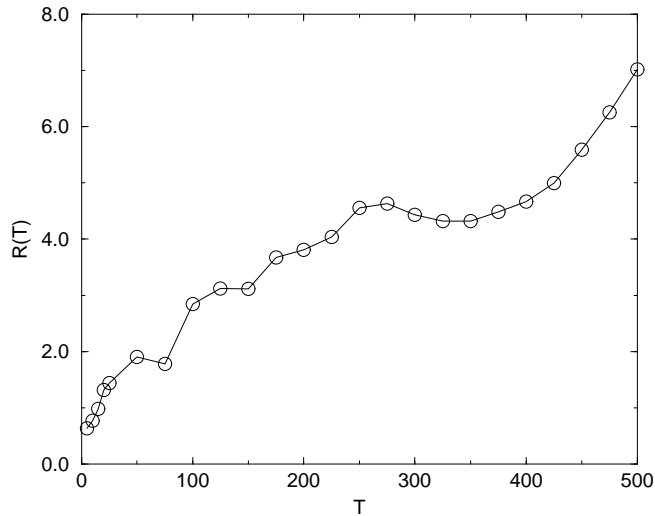,height=7.0cm,angle=0}}
\caption{Dispersion index $R(T)$ as a function of geological time $T$, 
as defined in Eq. \protect\ref{eq:dispersion}.}
\label{fig:dispersion}
\end{figure}

We measure as a function of $T$
the ratio $R(T)$ between the variance and the mean value of $S$
\be 
R(T)=\ov{\l\la S^2\r\ra-\l\la S\r\ra^2\over\l\la S\r\ra}\; , 
\label{eq:dispersion}
\ee
where the angular brackets denote the average respect to the 8 species
in our population and the overline denotes an average respect to 1000
extractions of the Poissonian variables. The resulting curve is shown
in Fig. \ref{fig:dispersion}.

If we assume that the mutations are mainly due to errors in the
replication accuracy, we should consider that $\mu=\mu_i$ depends on the
duration of a generation for species $i$ (in particular, it should be
inversely proportional). This is the generation time effect, that has
been shown to be present in real evolutionary data (Otha, 1993)
and to be stronger for synonymous mutations (for which $x=1$) than for
non-synonymous mutations, which are the subject of our study. We do not
consider here this effect, essentially for two reasons:

\begin{enumerate}
\item The mutation rate should also increase with the number of
  mitosis preceding reproduction, and this number is
  larger the larger the generation time. Thus the generation time
  effect is reduced in many cases.

\item It was shown that the dispersion index $R(T)$
  is significantly larger than unity even when the generation time effect
  and other lineage-depending effects are taken into account 
  (Gillespie, 1991). Gillespie named ``residual dispersion index'' the
  quantity $R(T)$ computed removing all lineages effects. Considering
  the same rate of mutation for all ``species'', we aim to study
  residual effects, which are most critical with respect to the neutral
  theory.
\end{enumerate}

\vspace{.5cm}
A last point remains to be made. Since $x$ has strong fluctuations, it
follows that the sequences in the neutral sets are not equivalent: 
sequences with a large value of $x(\S)$ should be
advantageous because their offsprings suffer a smaller fraction of
genetic deaths. Thus it could be thought that the population localizes
in a region of sequence space where $x(\S)$ is large, so that the
evolution is not anymore neutral at the level of population
genetics. However, we think that such phenomenon cannot take place.
In fact, the selective advantage of sequences with large $x(\S)$
is proportional to the mutation rate $\mu$. But it is known, for
instance from the theory of the error threshold (Eigen {\it et al.}, 1989),
that there
is a minimum selective advantage $u(\mu)$, increasing with the mutation
rate, below which natural selection is not able to fix advantageous
genotypes. We expect that the selective advantage implied by a larger
$x(\S)$ is always below the error threshold.
Moreover, since $x(\S)$ is a rather correlated quantity in
sequence space, and since the alleles present in a finite
population should be related through few mutations, the effective
selective advantage, related to the difference between the $x(\S)$ of
the alleles in the population, should be very small.

This effect however could act as a kind of negative feedback, reducing
the effect of the variations of $x$ on the variation of the
substitution rate. Another small reduction, present even when $x$ is
constant in the population, is due to a small correction to Kimura's
formula (Eq. \ref{kim}). It was shown in (Bastolla \& Peliti, 1991)
that the substitution
rate of a protein in a large population of asexually reproducing
individuals, evolving in a sequence landscape with sharply distributed
$x(\S)$, is given by
\be 
\mu_{subs}=\mu x/(1-\d)\;, 
\ee
where $\d\approx \mu N(1-x)$ is the fraction of the population eliminated by
lethal mutations. The factor $1/(1-\d)$  is due to a normalization
condition: the larger is $\d$, the easier is for the individuals who
suffered a neutral mutation to spread their genome in the population. 
Thus the effect of a variation in $x$ on the mutation rate is,  in the case
of a population, smaller than in the case of a single reproductive
lineage, where $\mu_{subs}=\mu x$, and the dispersion index should be
consequently slightly smaller.

Despite of these caveats, we think that our results are applicable
also at the level of population genetics.

\section{Discussion}
\label{sec:discussion}

We studied neutral evolution in sequence space. 
The theory of neutral evolution states that the mutations that do not
affect the biological activity of the protein are much more frequent
than advantageous
mutations. In our model the latter are not represented: all mutations
that are not neutral are assumed to be lethal. Neutrality is tested
imposing the conservation of the tridimensional structure, of its
thermodynamic stability and its kinetic accessibility.
Two main messages emerge.

The first one is that large differences in the genotype ({\it viz} the sequence)
are compatible with conservation of the phenotype 
({\it viz} the native structure).
The set of sequences which fold onto the same structure 
and are connected through point mutations is extended to form
a vast network in sequence space.
Two typical sequences belonging to this set, even if they are
evolutionarily related, may have a degree of homology as low as that
of random sequences. Thus sequence similarity is not a necessary condition 
for two proteins being evolutionarily related. 

The second message is that neutral evolution can be very irregular.
We have shown that the fraction of neutral mutations 
is a strongly fluctuating quantity inside a neutral set.
As a consequence of this fact, the trapping time on a given sequence has a
very broad distribution.
This observation is to our opinion very interesting for the
neutralist-selectionist controversy. One of the objections moved to
Kimura's theory is that, since the substitution process is assumed to
be Poissonian, it predicts a dispersion index $R(T)=V_S(T)/E_S(T)=1$,
where $V_S(T)$ and $E_S(T)$ are respectively the variance and the
expectation value of the number of substitutions happened in a time
$T$. For most proteins, a value of $R(T)$ significantly larger than 1 is
observed, and the discrepancy cannot be attributed to the generation time
effect nor to other lineage effects (Gillespie, 1991). Several
modifications of the neutral theory have been proposed in order to
reconcile it with this observation. It is not our aim to review them
here. We just note that, without additional hypothesis and with a
model that takes into account only neutral and lethal mutations (thus
without considering neither positive natural selection nor slightly
deleterious mutations) we find a dispersion index significantly larger
than 1, in agreement with many real data (however for some proteins, many of
which are hormones, the dispersion index is too large to be accounted
by this kind of explanation). Our results support the phenomenological
model of fluctuating neutral space introduced by Takahata (Takahata, 1987),
where the rate of substitution is assumed to vary at random after a
fixed number of substitutions.

The model that we studied is a very simplified one, 
and many question are open for discussion.
We recall here the ones that we judge the most serious:

\begin{enumerate}

\item We simulated the evolution of only one target structure. It
  would be interesting to see how our results change by changing the
  structure, and which properties of the structure (for instance
  compactness, locality of interactions, etc.) are important to determine
  the neutral mutation rate. However, the small
  number of folds occurring in natural proteins (at most some thousands) could
  be the ones to which corresponds the largest number of sequences in
  sequence space (Finkelstein {\it et al.}, 1993). Therefore, 
  structures characterized by a
  large neutral set, even if they are not typical, could be the most
  interesting ones from the biological point of view.

\item The size of the sequences examined is small, so that there are
  only two core residues. Considering more core residues could impose
  more constraints on the evolution and reduce the rate of neutral
  evolution. It would thus be interesting to repeat the same study for
  longer sequences.

\item The simple model we used do not allow for any discussion about
  biological activity, which would impose further constraints on the
  residues taking part to the active site.

\item In our model of evolution we assume that the environment remains
  fairly constant, so that the native structure favored by natural
  selection does not change throughout the evolution. This hypothesis
  is not unreasonable if the protein examined is an enzyme
  performing some chemical activity, since cells possess a high
  homeostasis, {\it i.e.} they can maintain a stable chemico-physical
  internal environment despite large perturbations in the external
  environment. However, it is quite likely that some large ecological
  and climatic changes have been responsible for molecular
  substitutions for which neutral theories, and our model in
  particular, do not apply (Gillespie, 1991).

\item We use a lattice model of a protein. This is a gross
  oversimplification, which does not capture essential features of
  real proteins like, for instance, the existence of secondary
  structure. Moreover, lattice models (as any other existing model)
  cannot be used for structure prediction. However, it has been
  argued that many qualitative features of lattice models are in good
  agreement with properties of real proteins.

\item We consider only point mutations, and not insertions and
  deletions, which also played an important role in evolution.

\end{enumerate}

So, which of our results can be applied to real proteins and
which cannot?
In our opinion, the limitations of our simulation should not modify
the qualitative picture. The existence of neutral networks
and the variability of neutral mutation rates are robust features that
occur in our model and seem to occur also in real proteins.

These two features originate in our model of evolution by imposing 
a ``phenotypic threshold'', below which the biological activity of the
target conformation is lost. Even if the threshold is very severe (we
impose $q(\C^*,\C_0)=1$, where $\C^*$ is the target structure and
$\C_0$ is the ground state of the sequence tested), the resulting
neutral network percolates sequence space.

This result is supported by the studies of (Rost, 1997)
and of (Babadje {\it et al.}, 1997).
Rost (Rost, 1997) showed that two structurally homologous proteins
have on the average a sequence homology only slightly larger than two
randomly chosen sequences.
Babadje and coworkers (Babadje {\it et al.}, 1997) arrived to the same
conclusion using a fold recognition
algorithm (Bowie {\it et al.}, 1991; Casari \& Sippl, 1992)
which measures the compatibility between the target
fold and a new sequence. Their method has the advantage of
considering PDB proteins, but the disadvantage that fold recognition
techniques,  even if good success is occasionally scored in structure
prediction, do not guarantee to give the right answer.
This holds in particular in the case where the sequence is not a real protein,
which is known to have a unique stable fold and to have been 
the outcome of the evolutionary process (only a very small number of
folds occurred in evolution, and the success of fold recognition
methods is also due to this fact, while caution is needed when a
random sequence is studied).
Another difference between (Babadje {\it et al.}, 1997) and our work is that
there only mutations which
increase the distance from the starting sequence are
allowed. This is reminiscent of a zero temperature
Monte Carlo algorithm for the optimization of the Hamming distance
$d(\S(t),\S(0))$. This rule is biologically unrealistic, and we think
that it is the reason why the walks cannot reach 
the maximal distance. Despite of these points, it is very
encouraging that very different methods give qualitatively the same results
concerning the diffusion in sequence space
(unfortunately the authors of (Babadje {\it et al.}, 1997)
did not observe the fluctuations of $x$).

\section*{Acknowledgments}

We acknowledge interesting discussions with 
Peter Grassberger, Helge Frauenkron, Walter Nadler, Anna Tramontano,
Tim Gibson, Erich Bornberg-Bauer and Guido Modiano. This work was conceived
during the workshop on Protein
Folding organized at the ISI Foundation, Torino, Italy, February 9-13 1998.
Part of the work was made during the Euroconference on "Protein
Folding and Structure Prediction" organized at the ISI Foundation,
Torino, Italy, June 8-19, 1998. Computations were carried out at the HLRZ,
Forschungszentrum J\"ulich.

\end{document}